\documentclass[a4paper,9pt, onecolumn]{article}
\usepackage{lmodern}
\usepackage[utf8]{inputenc}
\usepackage[T1]{fontenc}
\usepackage[english]{babel}
\usepackage[usenames,svgnames]{xcolor}
\usepackage{setspace}
\usepackage{endnotes}
\usepackage[labelfont=bf]{caption}
\usepackage{subcaption}
\usepackage{authblk}
\usepackage[pdftex]{graphicx} 
\usepackage[pdftex,pageanchor=true,hyperindex=true]{hyperref}
\usepackage{epstopdf} % to change .eps to .pdf
\usepackage{bookmark}
\usepackage{pdfpages}
\usepackage{array}
\usepackage{type1cm,soul}
\usepackage{tabularx}
\usepackage{ifthen} 
\usepackage{lscape}
\usepackage{amsmath}
\usepackage{amsfonts}
\usepackage{wasysym}
\usepackage{fullpage}
\usepackage{xstring}
\usepackage{titlesec}
\usepackage{colortbl}
\usepackage{tcolorbox}
\usepackage{calc}
\usepackage{enumitem}
\usepackage{gensymb} %for the degree symbol
\usepackage{pifont} %for checkmarks
\usepackage{booktabs} %for the tables
\usepackage{multirow} %for the tables
\usepackage{bm} %for bold math
\usepackage{siunitx}

\graphicspath{{Figures/}}

\font\titlefont=cmr12 at 15pt
\title{\titlefont Millisecond speed deep learning based proton dose calculation with Monte Carlo accuracy}

\author[]{Oscar Pastor-Serrano}
\author[]{Zoltán Perkó}

\affil[]{Delft University of Technology,\\ Department of Radiation Science and Technology, Delft, Netherlands}
\date{}

\begin{document}
\newcommand{\FixRef}[3][sec:]
{\IfBeginWith{#2}{#3}
	{\StrBehind{#2}{#3}[\RefResult]}
	{\def\RefResult{#2}}\IfBeginWith{#1}{#3}
	{\StrBehind{#1}{#3}[\RefResultb]}
	{\def\RefResultb{#1}}}

\newcommand{\secref}[1]
{\FixRef{#1}{sec:}Section~\ref{sec:\RefResult}}
\newcommand{\secreff}[1]
{\FixRef{#1}{sec:}in Section~\ref{sec:\RefResult}}
\newcommand{\Secreff}[1]
{\FixRef{#1}{sec:}In Section~\ref{sec:\RefResult}}
\newcommand{\secrefm}[2]
{\FixRef[#2]{#1}{sec:}Sections~\ref{sec:\RefResult}-\ref{sec:\RefResultb}}
\newcommand{\secreffm}[2]
{\FixRef[#2]{#1}{sec:}in Sections~\ref{sec:\RefResult}-\ref{sec:\RefResultb}}
\newcommand{\Secreffm}[2]
{\FixRef[#2]{#1}{sec:}In Sections~\ref{sec:\RefResult}-\ref{sec:\RefResultb}}
\newcommand{\figref}[1]
{\FixRef{#1}{fig:}Figure~\ref{fig:\RefResult}}
\newcommand{\figrefm}[2]
{\FixRef[#2]{#1}{fig:}Figures~\ref{fig:\RefResult}-\ref{fig:\RefResultb}}
\newcommand{\figreff}[1]
{\FixRef{#1}{fig:}in Figure~\ref{fig:\RefResult}}
\newcommand{\figreffm}[2
]{\FixRef[#2]{#1}{fig:}in Figures~\ref{fig:\RefResult}-\ref{fig:\RefResultb}}
\newcommand{\Figreff}[1]
{\FixRef{#1}{fig:}In Figure~\ref{fig:\RefResult}}
\newcommand{\Figreffm}[2]
{\FixRef[#2]{#1}{fig:}In Figures~\ref{fig:\RefResult}-\ref{fig:\RefResultb}}
\newcommand{\tabref}[1]
{\FixRef{#1}{tab:}Table~\ref{tab:\RefResult}}
\newcommand{\tabreff}[1]
{\FixRef{#1}{tab:}in Table~\ref{tab:\RefResult}}
\newcommand{\Tabreff}[1]
{\FixRef{#1}{tab:}In Table~\ref{tab:\RefResult}}
\newcommand{\tabrefm}[2]
{\FixRef[#2]{#1}{tab:}Tables~\ref{tab:\RefResult}-\ref{tab:\RefResultb}}
\newcommand{\tabreffm}[2]
{\FixRef[#2]{#1}{tab:}in Tables~\ref{tab:\RefResult}-\ref{tab:\RefResultb}}
\newcommand{\Tabreffm}[2]
{\FixRef[#2]{#1}{tab:}In Tables~\ref{tab:\RefResult}-\ref{tab:\RefResultb}}
\newcommand{\egyref}[1]
{\FixRef{#1}{eq:}Equation~\ref{eq:\RefResult}}
\newcommand{\eqreff}[1]
{\FixRef{#1}{eq:}in Equation~\ref{eq:\RefResult}}
\newcommand{\Eqreff}[1]
{\FixRef{#1}{eq:}In Equation~\ref{eq:\RefResult}}
\newcommand{\eqrefm}[2]
{Equations~\ref{eq:#1}-\ref{eq:#2}}
\newcommand{\eqreffm}[2]
{\FixRef[#2]{#1}{eq:}in Equations~\ref{eq:\RefResult}-\ref{eq:\RefResultb}}
\newcommand{\Eqreffm}[2]
{\FixRef[#2]{#1}{eq:}In Equations~\ref{eq:\RefResult}-\ref{eq:\RefResultb}}
\newcommand{\charef}[1]
{\FixRef{#1}{cha:}Chapter~\ref{cha:\RefResult}}
\newcommand{\chareff}[1]
{\FixRef{#1}{cha:}in Chapter~\ref{cha:\RefResult}}
\newcommand{\Chareff}[1]
{\FixRef{#1}{cha:}In Chapter~\ref{cha:\RefResult}}

\newcommand*{\dd}{\mathrm{d}}
\newcommand{\ui}[1]{\textit{\textbf{#1}}}
\newcommand{\mx}[1]{\underline{\underline{#1}}}
\newcommand{\diff}[2]{\dfrac{\dd #1}{\dd #2}}
\newcommand{\pdiff}[2]{\dfrac{\partial #1}{\partial #2}}
\newcommand{\dhl}{\hline\hline}
\newcommand{\rb}[1]{\left(#1\right)}
\newcommand{\sqb}[1]{\left[#1\right]}
\newcommand{\tb}[1]{\left<#1\right>}
\newcommand{\cb}[1]{\left\{#1\right\}}
\newcommand{\abs}[1]{\left|#1\right|}
\newcommand{\dspm}[1]{\begin{displaymath}#1\end{displaymath}}
\newcommand{\ds}{\displaystyle}
\newcommand{\pow}[2]{\cdot #1^{#2}}
\newcommand{\evat}[2]{\left.#1\right|_{#2}}
\newcommand{\ifrac}[2]{\ds #1 / #2}
\newcommand{\ab}{\ifrac{\alpha}{\beta}}
\newcommand{\BED}{\text{BED}}
\newcommand{\norm}[1]{\left\lVert#1\right\rVert}

\newcommand{\cmark}{\ding{51}}%
\newcommand{\xmark}{\ding{55}}%

\def\thetable{\Roman{table}}
\thispagestyle{empty}
\onecolumn
\maketitle
\noindent

\begin{abstract}
Next generation online and real-time adaptive radiotherapy workflows require precise particle transport simulations in sub-second times, which is unfeasible with current analytical pencil beam algorithms (PBA) or stochastic Monte Carlo (MC) methods. We present a deep learning based millisecond speed dose calculation algorithm (DoTA) accurately predicting the dose deposited by mono-energetic proton pencil beams for arbitrary energies and patient geometries. Given the forward-scattering nature of protons, we frame 3D particle transport as modeling a sequence of 2D geometries in the beam's eye view. DoTA combines convolutional neural networks extracting spatial features (e.g., tissue and density contrasts) with a transformer self-attention backbone that routes information between the sequence of geometry slices and a vector representing the beam’s energy, and is trained to predict low noise MC simulations of proton beamlets using 80,000 different head and neck, lung, and prostate geometries. Predicting beamlet doses in $5\pm4.9$ ms with a very high gamma pass rate of $99.37\pm1.17$\% (1\%, 3 mm) compared to the ground truth MC calculations, DoTA significantly improves upon analytical pencil beam algorithms both in precision and speed. Offering MC accuracy 100 times faster than PBAs for pencil beams, our model calculates full treatment plan doses in \SIrange{10}{15}{\second} depending on the number of beamlets (800-2200 in our plans), achieving a $99.70\pm0.14$\% (2\%, 2 mm) gamma pass rate across 9 test patients. Outperforming all previous analytical pencil beam and deep learning based approaches, DoTA represents a new state of the art in data-driven dose calculation and can directly compete with the speed of even commercial GPU MC approaches. Providing the sub-second speed required for adaptive treatments, straightforward implementations could offer similar benefits to other steps of the radiotherapy workflow or other modalities such as helium or carbon treatments.
\end{abstract}

\section{Introduction}
\label{sec:Introduction}
Radiotherapy (RT) treatments intimately rely on accurate particle transport calculations. In Computed Tomography (CT) image acquisition \cite{Pereira2014} simulations of the interaction between photons, tissues and detectors are used to obtain a detailed 3D image of the patient anatomy, which can be delineated to localize target structures and organs-at-risk. Modern intensity modulated treatments \cite{Hussein2018,Meyer2018} require particle transport to compute the spatial distribution of physical dose delivered by thousands of individual electron, photon, proton or other heavy ion beamlets (aimed at the patient from a few different beam angles), based on which the beamlet intensities can be optimized. Treatment plans -- especially sensitive proton and ion treatments -- must also be repeatedly evaluated under uncertainties (e.g., setup and range errors, tumor motion or complex anatomical changes) to ensure sufficient plan robustness, requiring recalculating the dose distribution in many different scenarios \cite{Perko2016,vanderVoort2016,RojoSantiago2021}. With RT practice steadily moving towards adaptive treatments, accurate, fast and general purpose dose (and particle transport) calculations represent an increasingly pressing, currently unmet need in most clinical settings.

We focus our attention specifically to proton dose calculations due to their more challenging nature caused by higher sensitivity and complexity compared to traditional photons. Current physics-based tools -- by and large falling into 2 categories: analytical pencil beam algorithms (PBAs) \cite{Hong1996,Schaffner1999} and stochastic Monte Carlo (MC) simulations --  offer a trade-off between speed and precision. While PBAs yield results without the computational burden of MC engines, their accuracy is severely compromised in highly heterogeneous or complex geometries, making slow and clinically often not affordable MC approaches necessary \cite{Teoh2020, Schuemann2015, Taylor2017,Grassberger2014,Saini2017}. The problem is most acute for online (and ultimately real-time) adaptive proton therapy aiming at treatment correction prior to (or even during) delivery to account for inter-fractional anatomical changes, motion due to breathing, coughs or intestinal movements. To become reality, such adaptive treatments require algorithms yielding MC accuracy with sub-second speed.

Reducing dose calculation times is an active area of research, with most works focusing on improving existing physics-based algorithms or developing deep learning frameworks. Several studies benefit from the parallelization capabilities of Graphics Processing Units (GPUs) to massively speed up MC simulations, reducing calculations times down to the range of few seconds \cite{Fracchiolla2021, Tseung2015} to minutes \cite{Ma2014, Gajewski2021,Pepin2018,Wang2016,Qin2016}, with simulation speeds up to $10^7$ protons/s. Deep learning methods have also improved dose calculation times in several steps of the RT workflow \cite{Meyer2018}, although usually paying the price of limited versatility and generalization capabilities. Some initial studies apply variants of U-net \cite{Ronneberger2015} and Generative Adversarial Networks \cite{Goodfellow2014} to aid treatment planning by approximating dose distributions from 'optimal' plans in very specific scenarios based on historical data. As input to these convolutional architectures, most works use organ and tumor masks \cite{Chen2019,Fan2019, Nguyen2019,Kajikawa2019}, CT images \cite{Kearney2018} or manually encoded beam information \cite{Nguyen2019a,BarraganMontero2019} to directly predict full dose distributions, except for few papers predicting the required beam intensities needed to deliver such doses \cite{Lee2019,Wang2020}.

Regarding pure dose calculation, practically all deep learning applications rely on using computationally cheaper physics simulations as additional input apart from CTs. For photons, most works predict low noise MC dose distributions from high noise MC doses \cite{Peng2019, Peng2019a,Bai2021,Neph2021} or simple analytical particle transport calculations \cite{Xing2020a,Dong2020}, with some approaches also utilizing additional manually encoded beam/physics information such as fluence maps \cite{Fan2020,Xing2020,Zhu2020,Kontaxis2020,Tsekas2021}. For protons, we are only aware of 2 papers \cite{Wu2021,Javaid2021} that compute low noise MC proton dose distributions via deep learning, both using cheap physics models (noisy MC and PBA) as input. While providing significant speed-up compared to pure physics-based algorithms, some even reaching sub-second speeds, all these works depend on secondary physics models to produce their output or are trained to predict only full plan or field doses for specific treatment sites. As a result, these methods do not qualify as generic dose algorithms and do not generalize to other steps of the RT workflow outside their original scope, e.g., to different plan or field configurations, treatment sites, or applications needing the individual dose distribution from each beamlet separately (such as treatment adaptation). 

\begin{figure*}[t]
	\centering
	\includegraphics[width=0.99\textwidth]{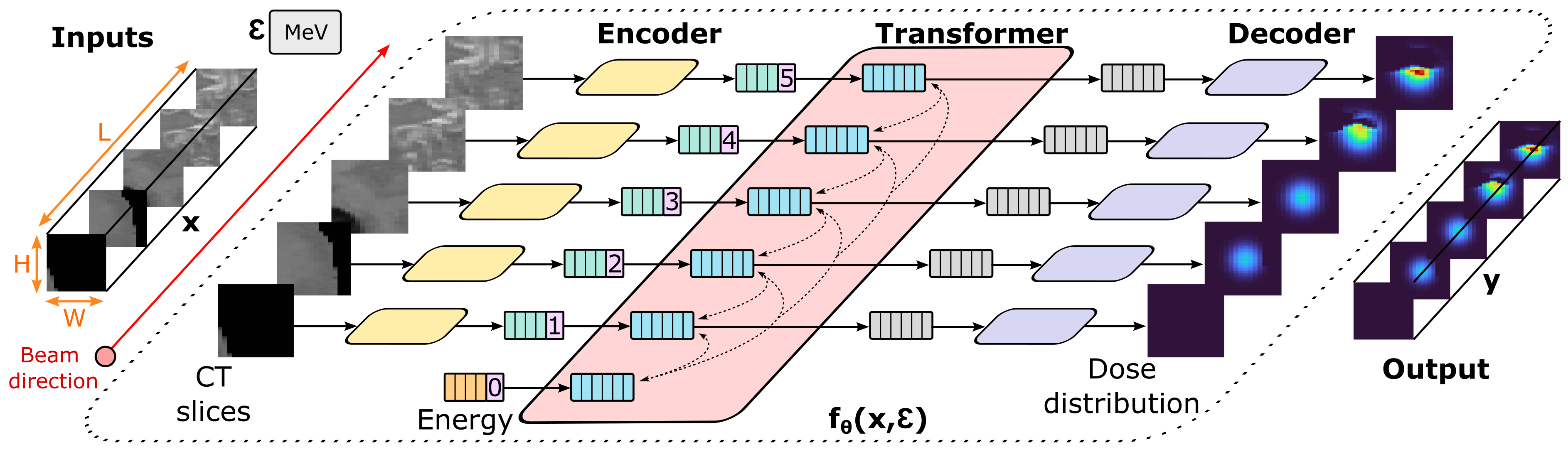} 
	\caption{\textbf{Dose transformer algorithm (DoTA).} A data-driven model learns a mapping $\bm{y} = f_{\bm{\theta}}(\bm{x}, \varepsilon)$ between input CT cubes $\bm{x}$ and energies $\varepsilon$ and output dose distributions $\bm{y}$. CT and dose distribution 3D volumes are both treated as a sequence of 2D slices in the beam's eye view. An encoder and a decoder individually transform each 2D slice into a feature vector and vice versa, whereas a transformer backbone routes information between different vectors along beam depth.}
	\label{fig:Model}
\end{figure*}

Instead, our study focuses on learning particle transport physics to substitute generic proton dose engines, providing millisecond speed and high accuracy, and is in principle applicable to all RT steps requiring dose calculations (e.g., dose-influence matrix calculation, dose accumulation, robustness evaluation). Our approach builds upon a previous study \cite{Neishabouri2021} using Long Short-term Memory (LSTM) networks \cite{Hochreiter1997} to sequentially calculate proton pencil beam dose distributions from relative stopping power slices in sub-second times, but with the major disadvantage of requiring a separate model per beam energy. As shown in \figref{Model}, we frame proton transport as modeling a sequence of 2D geometry slices in the beam's eye view, introducing an attention-based Transformer backbone \cite{Vaswani2017} that dynamically routes information between elements of the sequence along beam depth. We extend on our previous work only focusing on lung cancer \cite{Pastor-Serrano2021}, training with a larger set of patients and treatment sites, and evaluating performance both for individual pencil beams and full treatment plans. The presented Dose Transformer Algorithm (DoTA) -- able to learn the physics of energy dependence in proton transport via a single model -- can predict low noise MC proton pencil beam dose distributions purely from beamlet energy and CT data in $\approx\SI{5}{\milli\second}$. Based on our experiments and available literature data, in terms of accuracy and overall speed DoTA significantly outperforms pencil beam algorithms and all other deep learning approaches (e.g., LSTM models \cite{Neishabouri2021} and 'denoising' networks \cite{Wu2021, Javaid2021}), representing the current state-of-the-art in data-driven proton dose calculations and directly competing with (and even improving on) GPU Monte Carlo approaches.

\section{Methods and materials}
\label{sec:Methods}
The problem of dose calculation is common to many steps of RT workflow and ultimately involves estimating the spatial distribution of physical dose from thousands of pencil beams. A generic deep learning dose engine must be capable of calculating 3D dose distributions for arbitrary patient geometries purely from a list of beam directions and energies for a given beam model, without being conditioned on the type of treatment or task being solved. Therefore, our objective is to accurately predict dose distributions $\bm{y}$ from individual proton beamlets in sub-second speed, given patient geometries $\bm{x}$ and beam energies $\varepsilon$. We introduce DoTA, a parametric model that implicitly captures particle transport physics from data and learns the function $\bm{y} = f_{\bm{\theta}}(\bm{x},\varepsilon)$ via a series of artificial neural networks with parameters $\bm{\theta}$. 
 
% Framing the problem as sequence modeling
In particular, DoTA learns a mapping between a 3D CT input voxel grid $\bm{x}\in\mathbb{R}^{L\times H\times W}$ and output dose distribution $\bm{y}\in\mathbb{R}^{L\times H\times W}$ conditioned on the energy $\varepsilon\in\mathbb{R}^+$, where $L$ is the depth (in the direction of beam propagation), $H$ is the height and $W$ is the width of the grid. While traditional physics-based calculation tools process the entire geometry, we crop and interpolate the CT to the reduced sub-volume seen by protons as they travel through the patient, with a fixed $\SI{2}{\milli\meter}\times\SI{2}{\milli\meter}\times\SI{2}{\milli\meter}$ resolution and $L\times H\times W$ size. Framing proton transport as sequence modeling, DoTA processes the input volume as a series of $L$ 2D slices in the forward beam direction. Ideally, the exchange of information between the different elements in the sequence should be dynamic, i.e, the contribution or impact of each 2D slice to the sequence depends on both its position and material composition. Unlike other types of artificial neural networks, the Transformer architecture \cite{Vaswani2017} --- and specifically the self-attention mechanism --- is notably well suited for this.

% Transformer and why it's better than LSTM
Recently, Transformer-based architectures have replaced their recurrent counterparts in many natural language processing \cite{Devlin2019, Brown2020} and computer vision tasks \cite{Ramachandran2019, Dosovitskiy2020, Touvron2020, DAscoli2021}. For modeling the sequentiality in proton transport physics, the advantage of Transformers with respect to LSTM frameworks is two-fold. First, every element can directly access information at any point in the sequence without requiring an internal hidden state,  which is crucial to include beam energy dependence. The routing of information --- referred to as self-attention --- is different for every element, allowing each geometry slice to be independently transformed based on the information it selectively gathers from other slices in the sequence. Second, Transformers allow manually encoding the mostly forward scattering nature of proton transport by restricting interaction to only previous slices via causal attention. Transformers typically run multiple self-attention operations in parallel (known as attention heads), with each head focusing on modeling separate features of the sequence. We provide a detailed description of the fundamentals of self-attention and the Transformer module in Appendix \ref{Appendix:attention}.

\subsection{Model architecture and training} 
\begin{figure*}[t]
	\centering
	\includegraphics[width=0.99\textwidth]{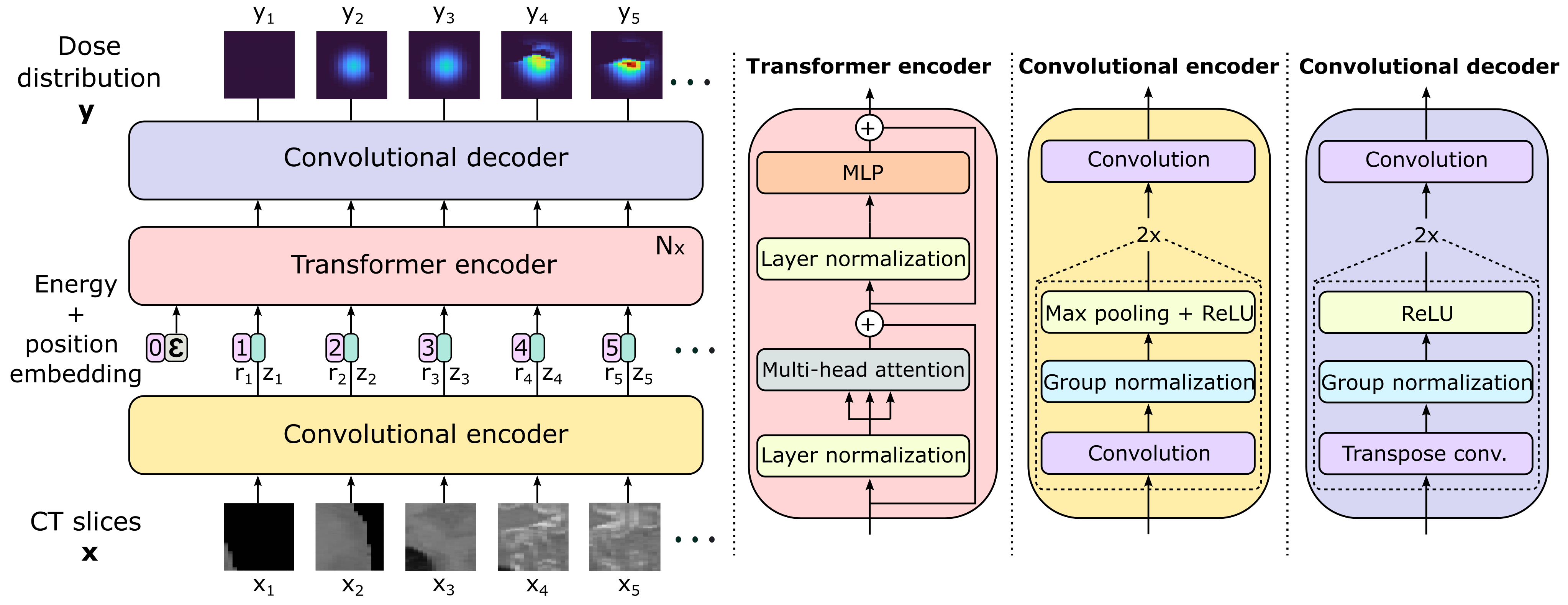} 
	\caption{\textbf{DoTA architecture}. We treat the input and output 3D volumes as a sequence of 2D slices. A convolutional encoder extracts important geometrical from each slice into a feature vector. The particle energy is added at the beginning of the resulting sequence. A transformer encoder with causal self-attention subsequently combines information from the different elements of the sequence. Finally, a convolutional decoder individually transforms the low-dimensional vectors into output 2D dose slices.}
	\label{fig:Architecture}
\end{figure*}

\figref{Architecture} shows DoTA's architecture, which first applies the same series of convolutions to each 2D slice of the input sequence $\{\bm{x}_i|\bm{x}_i\in\mathbb{R}^{1\times H\times W},\forall i=1,...,L \}$ separately. This convolutional encoder contains two blocks --- both with a convolution, a Group Normalization (GN) \cite{Wu2020} and a pooling layer, followed by a Rectified Linear Unit (ReLU) activation --- which extract important features from the input, e.g., material contrasts and tissue boundaries. After the second block, the outputs of a final convolution with $K$ filters are flattened into a vector of embedding dimension $D=H'\times W'\times K$, where $H'$ and $W'$ are the reduced height and width of the images after the pooling operations. The convolutional encoder applies the same operation to every element $\bm{x}_i$, resulting in a sequence of $L$ vectors $\{\bm{z}_i|\bm{z}_i\in\mathbb{R}^{D},\forall i=1,...,L \}$ referred to as tokens in the remainder of the paper.

A Transformer encoder models the interaction between tokens $\bm{z}_i$ via causal self-attention, resulting in an output sequence  $\bm{z}'\in\mathbb{R}^{D}$. Since Transformers operate on sets and by default do not account for the relative position of the slices in the sequence, we add a learnable positional encoding $\bm{r}_i\in\mathbb{R}^D$ to each token $\bm{z}_i$, e.g., $\bm{r}_1$ is always added to the token $\bm{z}_1$ from the first slice seen by the proton beam. The energy dependence is included via a 0\textsuperscript{th} token $\bm{z}_0=\bm{W}_0\varepsilon\in\mathbb{R}^D$ at the beginning of the sequence, where  $\bm{W}_0\in\mathbb{R}^{D\times 1}$ is a learned linear projection of the beam energy $\varepsilon$. We use the standard pre-Layer Normalization (LN) \cite{Ba2016} Transformer block \cite{Xiong2020}, alternating LN and residual connections with a self-attention operation and a feed-forward block with two fully-connected layers, Dropout \cite{Srivastava2014} and a Gaussian Error Linear Unit activation \cite{Hendrycks2016}. 

Finally, a convolutional decoder independently transforms every output token to a 2D slice of the same size as the input $\{\bm{y}_i|\bm{y}_i\in\mathbb{R}^{1\times H\times W},\forall i=1,...,L \}$. The decoder's structure is identical to that of its encoder counterpart, but substituting the down-sampling convolution + pooling operation in the with an up-sampling convolutional transpose layer.

\paragraph{Dataset} We train DoTA to predict low noise MC dose distributions calculated with MCsquare \cite{Souris2016}, obtained using a set of 30 CT scans from prostate, lung and head and neck (H\&N) cancer patients \cite{Aerts2014, Aerts2015, Clark2013} with 2 mm isotropic grid resolution. Given that proton beams have approximately 25 mm diameter and travel up to 300 mm through a small sub-volume of the CT, we crop blocks $\bm{x}\in\mathbb{R}^{150\times 24\times 24}$ covering a volume of approximately $48\times 48\times 300 $ $\text{mm}^3$. From each patient CT, we obtain $\approx 2,500$ of such blocks --- corresponding to beamlets being shot at different angles and positions --- by effectively rotating and linearly interpolating the CT scan in steps of $10\degree$ and by applying $10$ mm lateral shifts. 

For each block, we calculate 2 different dose distributions using $10^7$ primary particles to ensure MC noise values around 0.3\% and always below 0.5\%, zeroing out dose values below noise levels. Both dose distributions correspond to a randomly sampled beam energy between 70 and 220 MeV, with a 140 MeV cap in lung and H\&N geometries given the potential to overshoot the patient. As a result, we obtain $\approx 80,000$ individual CT block--dose distribution input--output pairs. This amount is further quadrupled by rotating the CT and dose blocks in steps of $90\degree$ around the beam direction axis, yielding a final training dataset consisting of $\approx 320,000$ samples, 10\% of which are used as a validation set to prevent overfitting.

Our evaluation is based on an independent test set of 18 additional patients unseen during training, equally split into prostate, H\&N and lung. Half of these patients (3 prostate, 3 H\&N and 3 lung) are used to compare beamlet dose distributions, with the other half serving to evaluate DoTA's performance in full plans. 

\paragraph{Training details} The model is trained end-to-end using Tensorflow \cite{Abadi2015}, with the LAMB optimizer \cite{You2019} and 8 samples per mini-batch, limited by the maximum internal memory of the Nvidia Tesla T4{\textregistered} GPU used during our experiments. We use a mean squared error loss function and a scheduled learning rate starting at $10^{-3}$ that is halved every 4 epochs, with a restart after 28 epochs. In total, we train the model for 56 epochs, saving the weights resulting in the lowest validation mean squared error. The best performing model consists of one transformer block with 16 heads and 12 convolutional filters in the last encoder layer, as obtained from a hyperparameter grid search evaluating the lowest validation loss across all possible combinations of transformer layers $N\in\{1,2,4\}$, convolutional filters $K\in\{8,10,12,16\}$ and attention heads $N_h\in\{8,12,16\}$. Given the two down-sampling pooling operations, the transformer processes tokens of dimension $D=H/4\times W/4\times K$, which in our case with initial height $H=24$, width $W=24$, and $K=12$ kernels results in $D=432$. 

\subsection{Model evaluation}
Using the ground truth MC dose distributions in the test set, we compare DoTA to several data-driven dose engines, including LSTM models \cite{Neishabouri2021}, and deep learning frameworks using noisy MC \cite{Javaid2021} and PBA \cite{Wu2021} doses as additional input. Since PBA is the analytical dose calculation method commonly used in the clinic and one of DoTA's competitors in terms of speed and accuracy, we include the PBA baseline from the open-source treatment planning software matRad \cite{Wieser2017} (\url{https://e0404.github.io/matRad/}).

\paragraph{Test set accuracy metrics} In our evaluation, the main mechanism to compare predictions to ground truth 3D dose distributions from the test set is the gamma analysis \cite{Low1998}, further explained in Appendix \ref{Appendix:gamma}. To reduce the gamma evaluation to a single number per sample, we report the gamma pass rate as the fraction of passed voxels over the total number of voxels. All calculations are based on the PyMedPhys gamma evaluation functions (available at \url{https://docs.pymedphys.com}).  

Additionally, the average relative error $\rho$ is used to explicitly compare dose differences between two beamlet dose distributions. Given the predicted output $\bm{y}$ and the ground truth dose distribution $\bm{\hat{y}}$ with $n_v = L\times H\times W$ voxels, the average relative error can be calculated as 

\begin{equation}
	 \rho = \frac{1}{n_v} \frac{\norm{\bm{\hat{y}}-\bm{y}}_{L_1}}{\max{\bm{\hat{y}}}}\times 100.
\end{equation}

\paragraph{Experiments} A generic data-driven dose engine must yield accurate predictions for both single beamlet and full plan dose distributions. To ensure DoTA's suitability for replacing conventional particle transport tools in dose prediction tasks, we assess its performance in two different settings:

\begin{itemize}[leftmargin=*]
	\item Individual beamlets. First, we evaluate the speed and accuracy in predicting single beamlet doses for 9 patients in the test set and compare gamma pass rate distributions and inference times of DoTA, the LSTM models and the PBA baseline. Given the $\SI{2}{\milli\meter}\times\SI{2}{\milli\meter}\times\SI{2}{\milli\meter}$ grid resolution, a gamma evaluation $\Gamma(3\text{ mm},1\%)$ using a distance-to-agreement criterion $\delta=3$ mm ensures a neighborhood search of at least one voxel, while a dose criterion $\Delta=1\%$ disregards any uncertainty due to MC noise. Since DoTA's outputs are hardly ever 0 due to numerical inaccuracies of the last convolutional linear layer, and to disregard voxels not receiving any dose, we exclude voxels with doses below 0.1\% of the maximum dose for the gamma pass rate calculations, resulting in a stricter metric (as the many voxels with near 0 dose could artificially increase the passing rate). Additionally, we compute the relative error $\rho$ between PBA/DoTA predictions and MC dose distributions. For both  $\rho$ and the gamma pass rate, we compare probability densities across all test samples.
	\item Full plans. A treatment plan with 2 fields is obtained for the remaining 9 test set patients using matRad. Given the list of beam intensities and energies in the plan, we recalculate dose distributions using PBA, MCsquare \cite{Souris2016} and DoTA, and evaluate their performance via the gamma pass rate, masking voxels receiving a dose lower than 10\% of the maximum dose. For each field angle in the treatment plan, we rotate the original CT, calculate the dose from each beamlet and rotate back the entire field dose its original angle for dose accumulation. To allow for a fair comparison with other data-driven models --- referred to as baselines B1 \cite{Javaid2021} and B2 \cite{Wu2021} --- we compute three gamma evaluations  $\Gamma(1\text{ mm},1\%)$,  $\Gamma(2\text{ mm},2\%)$ and  $\Gamma(3\text{ mm},3\%)$ and compare the pass rate results to the available values in baseline  studies. For more information about the experiments, \tabref{experiments} contains a description of the metrics and evaluation settings.
\end{itemize}

\begin{table}[]
	\centering
	\caption{\textbf{Overview of experiments.} Summary of the experiments, metrics and baselines used to evaluate DoTA's accuracy. $D_{\text{max}}$ refers to the maximum dose value in a dose distribution and only voxels receiving dose above the cutoff level are included in the $\Gamma$ calculations.}
	\begin{tabular}{@{}lcccc@{}}
		\toprule
		\textbf{Experiment} & \textbf{Test data} & \multicolumn{1}{c}{\textbf{Metric}} & \multicolumn{1}{c}{\textbf{Dose cutoff (Gy)}} & \multicolumn{1}{c}{\textbf{Baseline}} \\ \midrule
		\multirow{3}{*}{Individual beamlets} & \multirow{3}{*}{3,888 pencil beams} & \multicolumn{1}{c}{\multirow{2}{*}{$\Gamma(3\text{ mm}, 1\%)$}} & 0 & LSTM \\
		&  & \multicolumn{1}{c}{} & 0.1\% of $D_{\text{max}}$ & PBA \\
		&  & Error $\rho$ & 0 & PBA \\ \midrule
		\multirow{2}{*}{Full plans} & \multirow{2}{*}{9 treatment plans} & $\Gamma(1\text{ mm}, 1\%)$ & 10\% of $D_{\text{max}}$ & PBA, B2 \\
		&  & $\Gamma(2\text{ mm}, 2\%)$ & 10\% of $D_{\text{max}}$ & B1 \\ \bottomrule
	\end{tabular}
	\label{tab:experiments}
\end{table}

\section{Results}
\label{sec:Results}
In this section, DoTA's performance and speed is compared to state-of-the-art models and clinically used methods. The analysis is three-fold: we assess the accuracy in predicting beamlet dose distributions and full dose distributions from treatment plans, and explore DoTAs' potential as a fast dose engine by evaluating its calculation runtimes. 

\subsection{Individual beamlets} 
\begin{figure}[]
	\begin{minipage}[t]{0.49\textwidth}
		\centering
		\includegraphics[width=\textwidth]{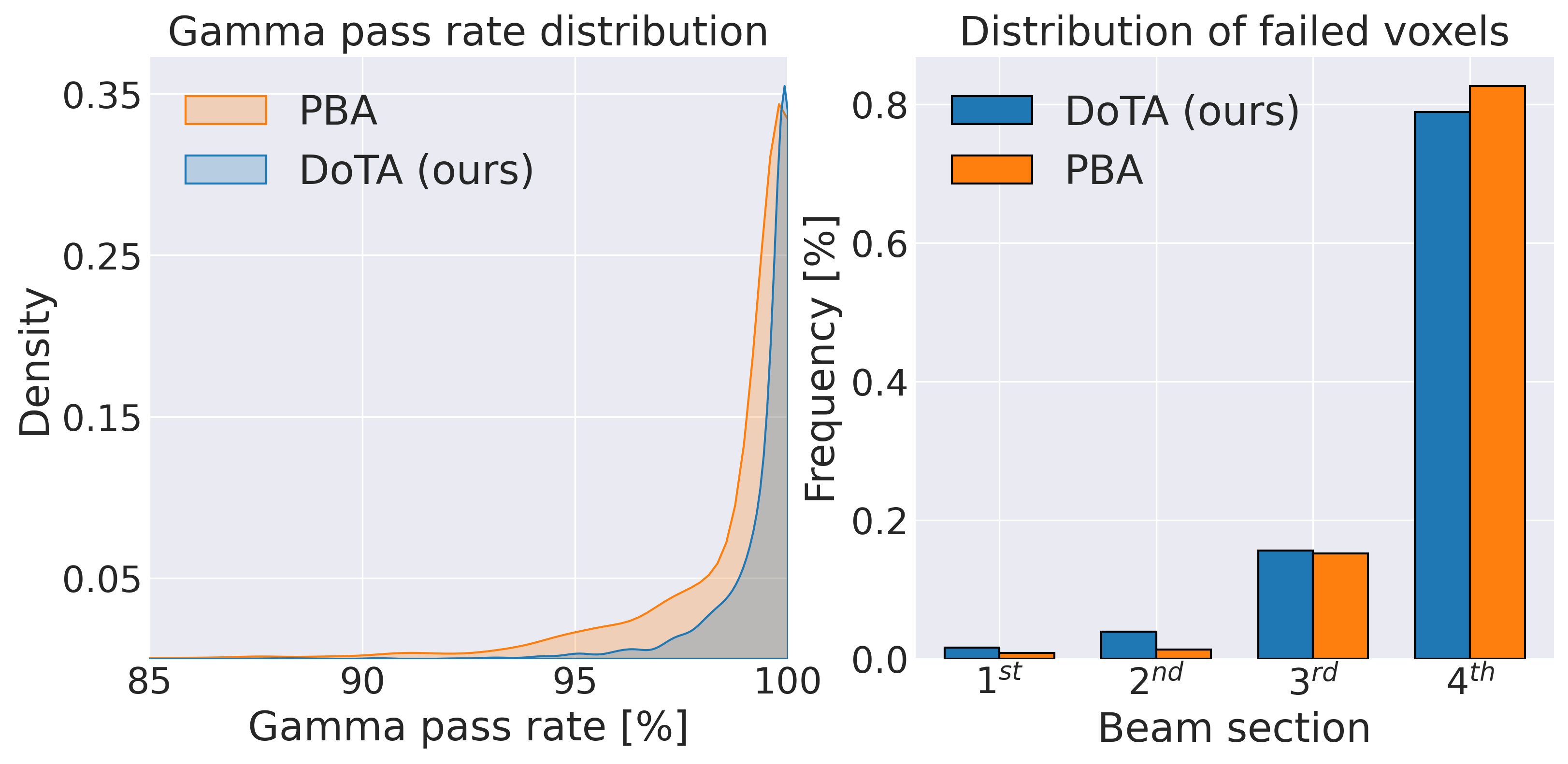}
		\captionof{figure}{\textbf{Gamma pass rate distribution.} (Left) Distribution of the gamma pass rates $\Gamma(3 \text{ mm},1\%)$ of the test samples for the pencil beam algorithm (PBA) and the presented DoTA model. (Right) Distribution of the failed voxels along the beam, where each bin is a section of the beam from dose entrance (1$^{\text{st}}$) to Bragg Peak and dose falloff (4$^{\text{th}}$). Each bin shows the ratio of the number of test set voxels that fail the gamma evaluation within a section divided by the total number of failed voxels.}
		\label{fig:gpr}
	\end{minipage}
	\hfill
	\begin{minipage}[t]{0.49\textwidth}
		\centering
		\includegraphics[width=\textwidth]{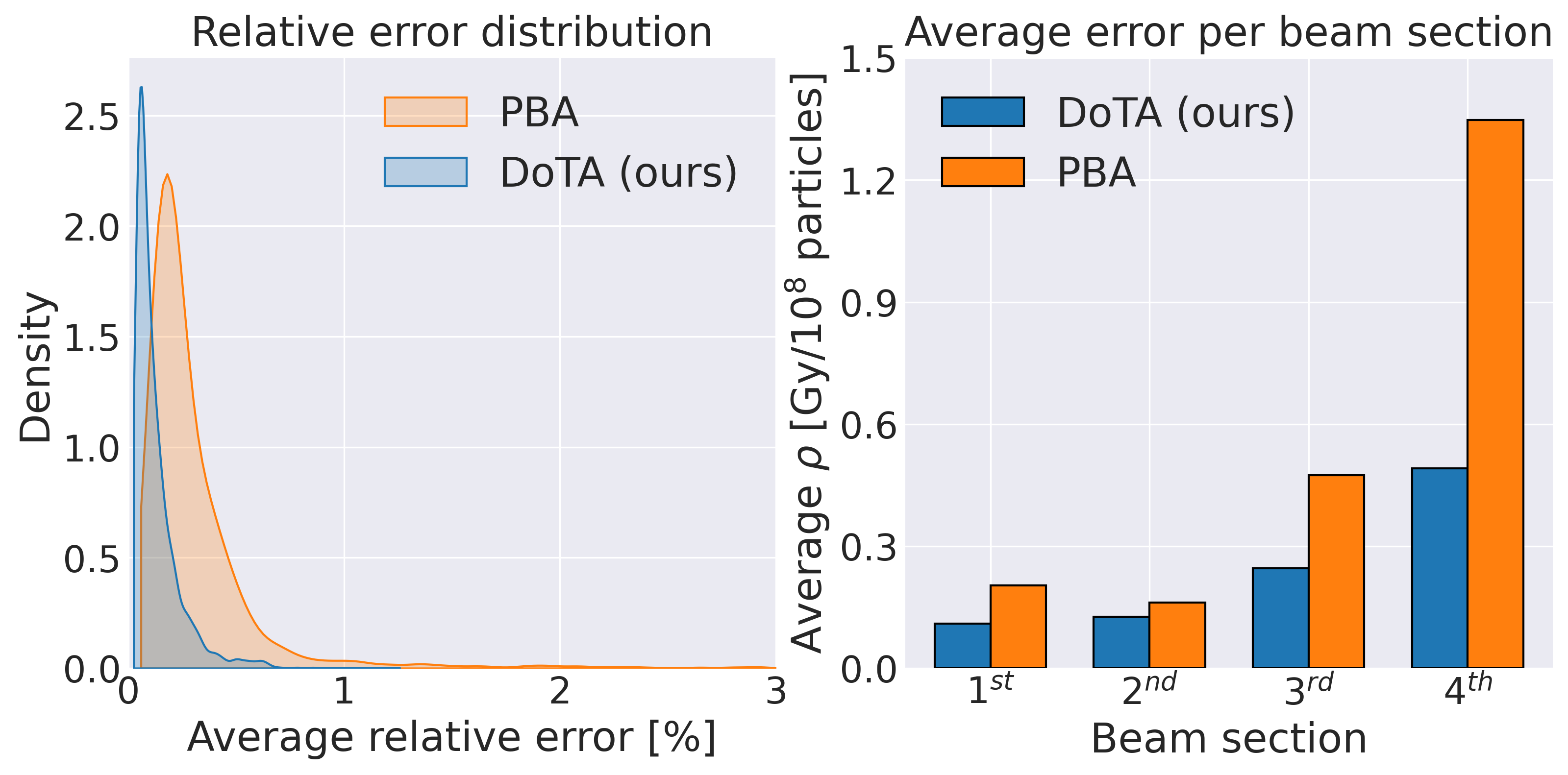}
		\captionof{figure}{\textbf{Average relative error $\rho$ distribution.} (Left) Distribution of the average relative error across the test samples for the pencil beam algorithm (PBA) and the presented DoTA model. (Right) Average relative error per beam section, where each bin is a section of the beam from dose entrance (1$^{st}$) to Bragg Peak and dose falloff (4$^{th}$). Each bin shows the average of the relative error values recorded within a section of the beam.}
		\label{fig:err}
	\end{minipage}
\end{figure}

For each individual beamlet in the test set, DoTA's predictions are compared to MC ground truth dose distributions using a $\Gamma(3 \text{ mm},1\%)$ gamma analysis. In \tabref{gpr}, we report the average, standard deviation, minimum and maximum of the distribution of gamma pass rates across test samples. By disregarding voxels whose dose is below 0.1\% of the maximum dose, our gamma evaluation approach is stricter than that of previous state-of-the-art studies \cite{Neishabouri2021}, where only voxels with a gamma value of 0 --- which typically correspond to voxels not receiving any dose --- are excluded from the pass rate calculation. Even with the stricter setting and including energy dependence, DoTA outperforms both the LSTM and PBA dose engines in all aspects: the average pass rates are higher, the standard deviation is lower, and the minimum is at least 5.5\% higher. The left plot in \figref{gpr} further demonstrates DoTA's superiority, showing a gamma pass rate distribution that is more concentrated towards higher values. The right plot in \figref{gpr} shows the proportion of voxels failing the gamma evaluation in each beam section, out of the total number of failed voxels, indicating for both PBA and DoTA that most of the failing voxels belong to the 4$^{\text{th}}$ section, i.e., the high energy region around the Bragg peak where the effect of tissue heterogeneity is most evident.

\begin{table}[]
	\centering
	\caption{\textbf{Gamma pass rate of beamlet dose distributions.} Gamma analysis results $\Gamma(3 \text{mm}, 1\%)$ for the presented DoTA, the pencil beam algorithm (PBA) and the LSTM models are listed. Gamma pass rates are calculated using test samples, with LSTM rates directly obtained from \cite{Neishabouri2021}. The reported values include the mean, standard deviation (Std), minimum (Min) and maximum (Max) across the test set for different treatment sites, and 'Multi-site' refers to computing statistics using all sites.}
	\begin{tabular}{@{}lcccccc@{}}
		\toprule
		\textbf{Model} & \textbf{Site} & \textbf{Energy (MeV)} & \textbf{Mean (\%)} & \textbf{Std (\%)} & \textbf{Min (\%)} & \textbf{Max (\%)} \\ \midrule
		\multirow{3}{*}{LSTM \cite{Neishabouri2021}} & \multirow{3}{*}{Lung} & 67.85 & 98.56 & 1.3 & 95.35 & 99.79 \\
		&  & 104.25 & 97.74 & 1.48 & 92.57 & 99.74 \\
		&  & 134.68 & 94.51 & 2.99 & 85.37 & 99.02 \\ \midrule
		\multirow{3}{*}{DoTA (ours)} & Lung & [70-220] & 99.46 & 0.81 & 93.19 & 100 \\
		& H\&N & [70-220] & 99.21 & 1.23 & 93.49 & 100 \\
		& Prostate & [70-220] & 99.51 & 1.46 & 94.06 & 100 \\ \midrule
		DoTA (ours) & Multi-site & [70-220] & 99.37 & 1.17 & 93.19 & 100 \\
		PBA \cite{Wieser2017} & Multi-site & [70-220] & 98.68 & 3.14 & 87.53 & 100 \\ \bottomrule
	\end{tabular}
	\label{tab:gpr}
\end{table}

\begin{table}[]
	\centering
	\caption{\textbf{Average relative error of beamlet dose distributions.} The reported values include the mean, standard deviation (Std), minimum (Min) and maximum (Max) values of the error $\rho$ between predictions and reference MC dose distributions, for both the pencil beam algorithm (PBA) and DoTA.}
	\begin{tabular}{@{}lcccc@{}}
		\toprule
		\textbf{Model} & \textbf{Mean (\%)} & \textbf{Std (\%)} & \textbf{Min (\%)} & \textbf{Max (\%)} \\ \midrule
		\begin{tabular}[c]{@{}l@{}}DoTA (ours)\end{tabular} & 0.126 & 0.109 & 0.025 & 1.258 \\
		PBA \cite{Wieser2017} & 0.306 & 0.309 & 0.059 & 4.077 \\ \bottomrule
	\end{tabular}
	\label{tab:err}
\end{table}

As an additional measure of model performance, \tabref{err} shows the mean and standard deviation of the relative error $\rho$ between predictions and ground truth MC dose distributions in test set. The results confirm DoTA's improvement, with mean, maximum error and standard deviation less than half of PBA's. The left plot in \figref{err} displays the distribution of $\rho$ across all test samples, showing that values are smaller and closer to 0 for DoTA. As with the gamma pass rate, the beam is divided in 4 sections from entrance (1$^{\text{st}}$) to the Bragg peak (4$^{\text{th}}$), and the average relative error per section is shown in the right plot in \figref{err}. Although both models show a similar trend with errors increasing towards the beam's end, DoTA is on average twice better than PBA.

\begin{figure}[]
	\centering
	\begin{subfigure}[t]{0.47\textwidth}
		\centering
		\includegraphics[width=\textwidth]{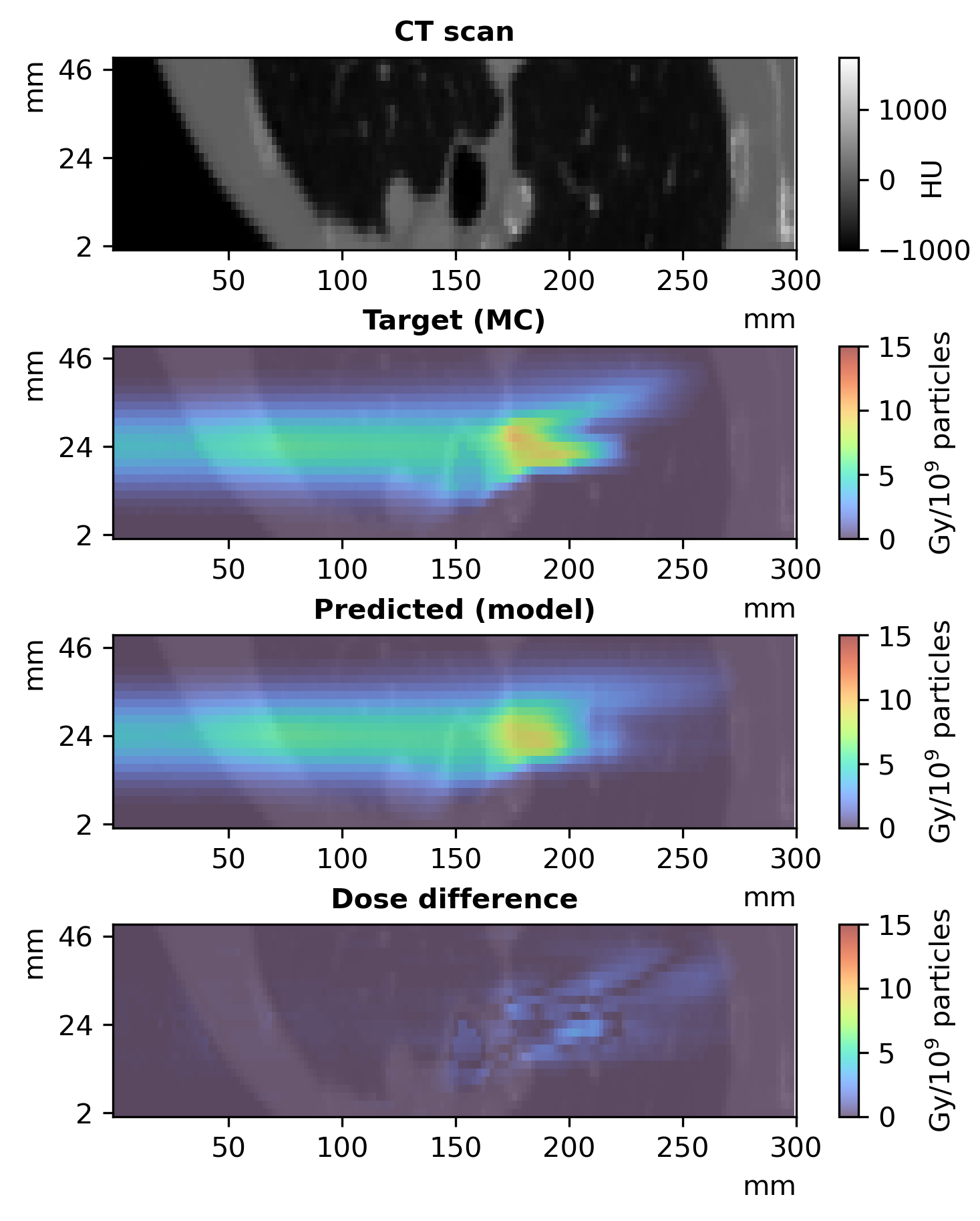}
		\caption{DoTA's worst prediction.}
		\label{fig:wdotal}
	\end{subfigure}
	\begin{subfigure}[t]{0.47\textwidth}
		\centering
		\includegraphics[width=\textwidth]{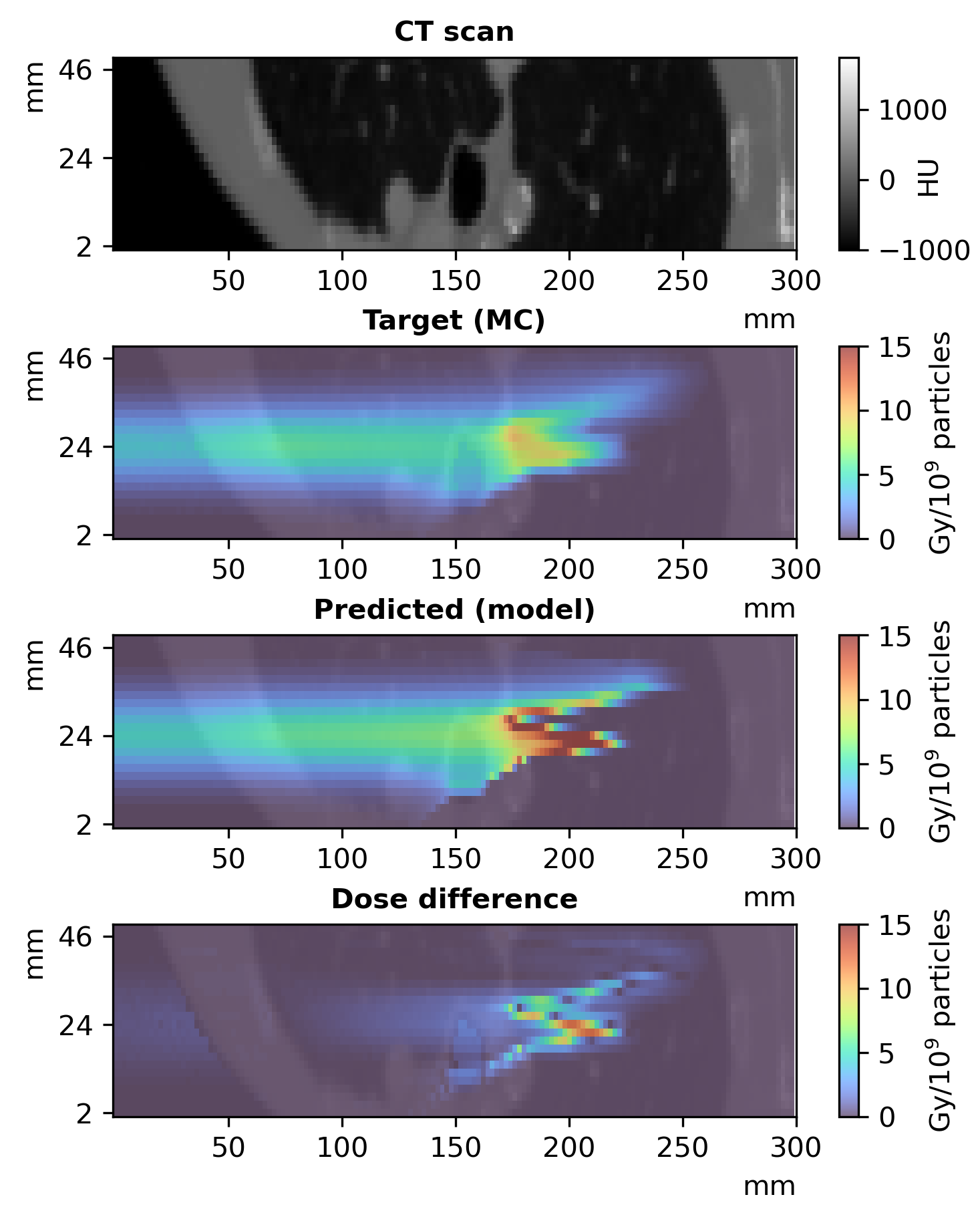}
		\caption{Pencil beam algorithm prediction of the worst DoTA sample.}
		\label{fig:wdotar}
	\end{subfigure}
	\begin{subfigure}[t]{0.47\textwidth}
		\centering
		\includegraphics[width=\textwidth]{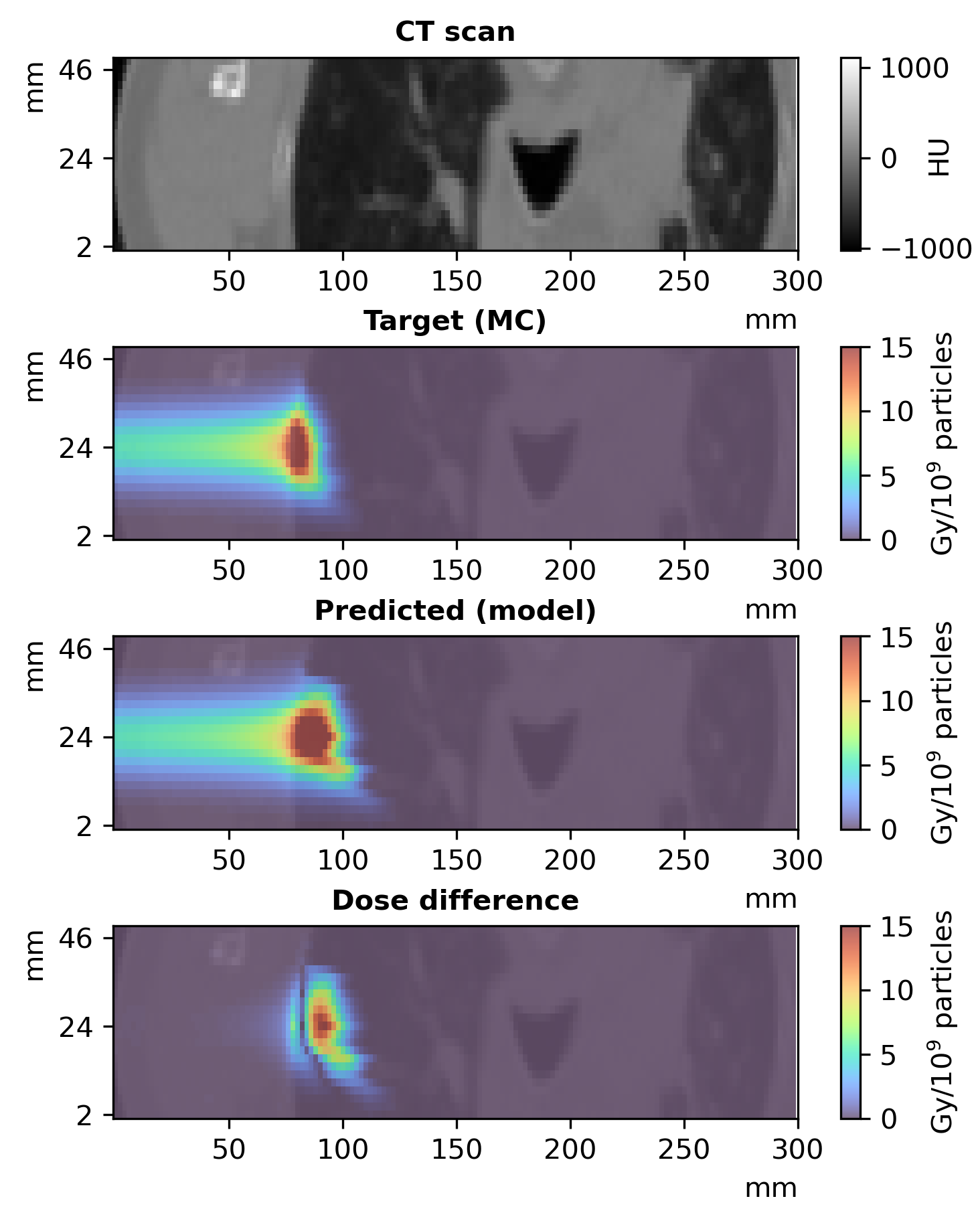}
		\caption{Pencil beam algorithm worst prediction.}
		\label{fig:wpbal}
	\end{subfigure}
	\begin{subfigure}[t]{0.47\textwidth}
		\centering
		\includegraphics[width=\textwidth]{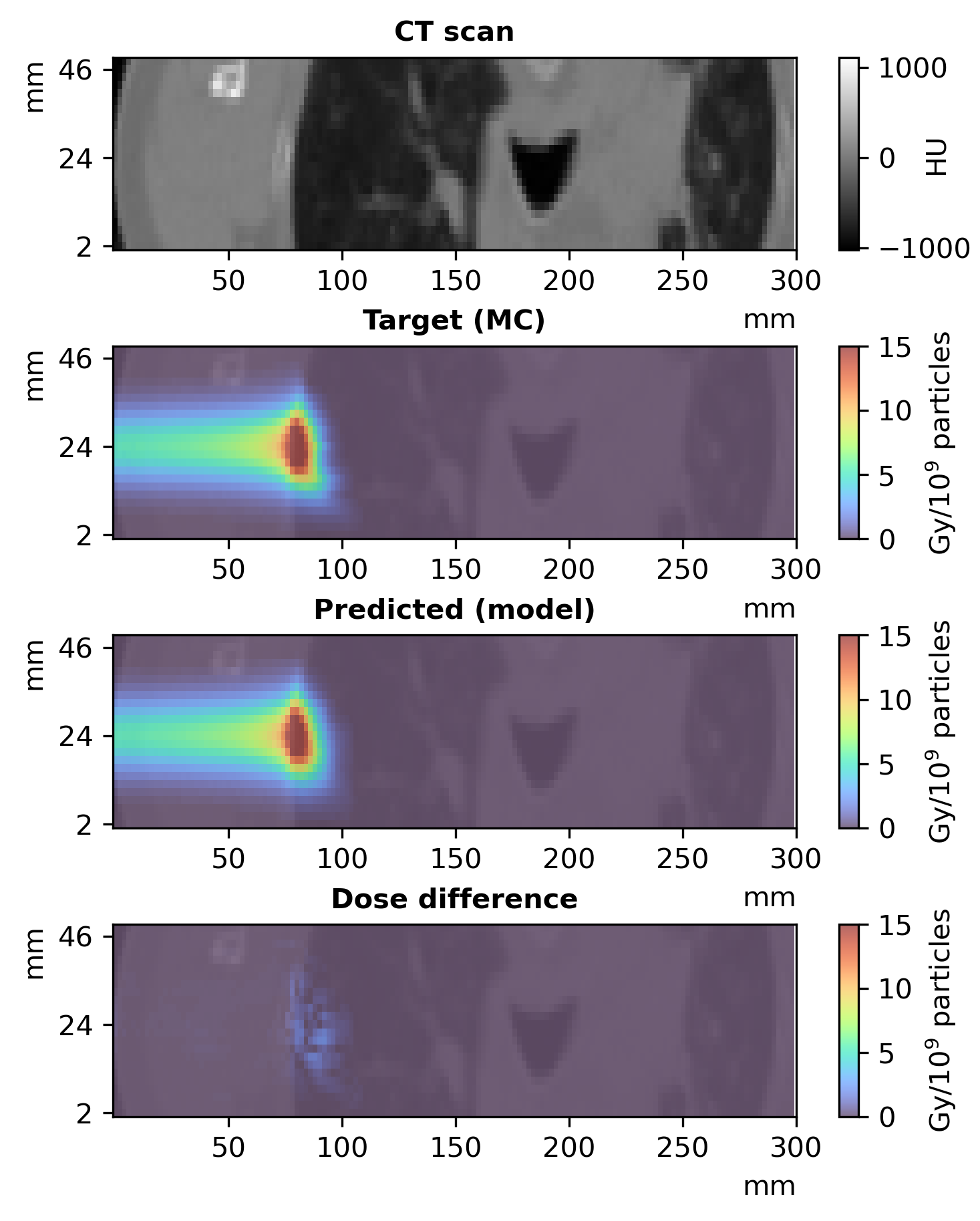}
		\caption{DoTA's prediction of the worst pencil beam algorithm sample.}
		\label{fig:wpbar}
	\end{subfigure}
	\caption{\textbf{Worst performing DoTA and PBA test sample.} (a) Worst performing test sample in the gamma evaluation for DoTA, with gamma pass rate of 93.19\%, and (b) the pencil beam algorithm (PBA) prediction for the same sample. (c) Worst performing prediction in the gamma evaluation across the test set for PBA, with gamma pass rate of 87.53\%, and  (d) DoTA's prediction of the same sample. In descending order, all 4 subplots show: the central slice of the 3D input CT grid, the MC ground truth dose distribution, the model's prediction and the dose difference between the predicted and MC beams.}
	\label{fig:worst}
\end{figure}

Finally, \figref{wdotar} shows DoTA's test sample with the lowest gamma pass rate, together with PBA's prediction of the same sample (\figref{wdotal}). Likewise, \figref{wpbal} and \figref{wpbar} show the predictions of the worst PBA sample from both models. In both cases, PBA results in errors as high as 80\% of the maximum dose, severely overdosing parts of the geometry, while for DoTA errors are below 20\% of the maximum dose.

%\begin{figure}[]
%	\centering
%	\begin{subfigure}[t]{0.40\textwidth}
%		\centering
%		\includegraphics[width=\textwidth]{Figures/wp_pba.png}
%		\caption{Pencil beam algorithm worst prediction}
%		\label{fig:worst_pba_l}
%	\end{subfigure}
%	\begin{subfigure}[t]{0.40\textwidth}
%		\centering
%		\includegraphics[width=\textwidth]{Figures/wp_dota.png}
%		\caption{DoTA's prediction of the worst pencil beam algorithm sample}
%		\label{fig:worst_pba_r}
%	\end{subfigure}
%	\caption{\textbf{Worst pencil beam algorithm (PBA) test sample.} (a) Worst performing prediction in the gamma evaluation across the test set for PBA, with gamma pass rate of 87.53\%, and  (b) DoTA's prediction of the same sample. Both plots displays the central slice of the 3D input CT grid, the MC ground truth dose distribution, the model prediction and the absolute dose difference between the predicted and MC beams.}
%	\label{fig:worst_pba}
%\end{figure}

\begin{table}[]
	\centering
	\caption{\textbf{Gamma pass rate of planned dose distributions.} Treatment plans of 9 test patients are recalculated using the presented DoTA model, and compared to ground truth MC dose distributions via 3 different gamma analysis: $\Gamma(1\text{ mm},1\%)$, $\Gamma(2\text{ mm},2\%)$ and $\Gamma(3\text{ mm},3\%)$. We additionally include the $\Gamma(1\text{ mm},1\%)$ pass rate for dose distributions recalculated by the pencil beam algorithm (PBA). The baseline B1 corresponds to a MC-denoising U-net \cite{Javaid2021}, while B2 is a U-net correcting PBA \cite{Wu2021}, whose values are directly taken for their corresponding papers. }
	\begin{tabular}{@{}lcccccccc@{}}
		\toprule
		\multirow{2}{*}{\textbf{Site}} & \multirow{2}{*}{\textbf{Patient}} & \multirow{2}{*}{\parbox{1.5cm}{\centering \textbf{Number of spots}}} & \multicolumn{3}{c}{\textbf{DoTA (ours)}} & \textbf{PBA}\cite{Wieser2017} & \textbf{B1} \cite{Javaid2021} & \textbf{B2} \cite{Wu2021} \\
		&  &  & $\Gamma(1,1\%)$ & $\Gamma(2,2\%)$ & $\Gamma(3,3\%)$ &  $\Gamma(1,1\%)$ & $\Gamma(2,2\%)$ & $\Gamma(1,1\%)$ \\ \midrule
		\multirow{3}{*}{Lung} & 1 & 954 & 95.86 & 99.73 & 99.99 & 80.38 & \multirow{3}{*}{84.1} & \multirow{3}{*}{89.7$\pm$3.8} \\
		& 2 & 2245 & 96.31 & 99.72 & 99.98 & 79.83 &  & \\
		& 3 & 1646 & 95.63 & 99.64 & 99.97 & 78.92 &  & \\ \midrule
		\multirow{3}{*}{H\&N} & 4 & 1554 & 95.02 & 99.39 & 99.81 & 68.32 & \multirow{3}{*}{76.5} & \multirow{3}{*}{92.8$\pm$2.9} \\
		& 5 & 1064 & 94.71 & 99.62 & 99.97 & 76.63 &  & \\
		& 6 & 708 & 96.93 & 99.88 & 99.99 & 83.02 &  & \\ \midrule
		\multirow{3}{*}{Prostate} & 7 & 1598 & 96.38 & 99.81 & 99.99 & 87.34 & \multirow{3}{*}{-} & \multirow{3}{*}{99.6$\pm$0.3} \\
		& 8 & 2281 & 95.78 & 99.82 & 99.99 & 77.12 &  &  \\
		& 9 & 1518 & 96.18 & 99.71 & 99.98 & 83.64 &  &  \\ \bottomrule
	\end{tabular}
	\label{tab:gprplans}
\end{table}

\subsection{Full dose recalculation} To assess the feasibility of using DoTA as a dose engine in real clinical settings, we recalculate full dose distributions from treatment plans and compare them to MC reference doses via 3 different gamma analysis: $\Gamma(1 \text{ mm},1\%)$, $\Gamma(2\text{ mm},2\%)$ and $\Gamma(3\text{ mm},3\%)$, in decreasing order of strictness. The resulting gamma pass rates for each of the 9 test patients are shown in \tabref{gprplans}, showing values that are consistently high and similar across treatment sites, always at least 10\% higher than PBA. We additionally compare DoTA to recently published state-of-the-art deep learning approaches: a MC-denoising U-net \cite{Javaid2021} (B1), and a U-net correcting PBA \cite{Wu2021} (B2). Except for the prostate plans, DoTA outperforms both approaches, even without requiring the additional physics-based input.

\subsection{Runtime} Apart from high prediction accuracy, fast inference is critically important for clinical applications. \tabref{time} displays the mean and standard deviation runtime taken by each model to predict a single beamlet. Being particularly well-suited for GPUs, DoTA is on average faster than LSTM and physics-based engines, offering more than 100 times speed-up with respect to PBA. Additionally, although dependent on hardware, DoTA approximates doses four orders of magnitude faster than MC, providing millisecond dose calculation times without requiring any extra computations for real-time adaptive treatments. 

Regarding full dose recalculation from treatment plans, Figure \ref{timeplans} shows total runtimes for DoTA using both GPU and CPU hardware, including all steps from loading CT and beamlet weights from plan data files, necessary CT rotations and interpolations, DoTA dose inference time and reverse rotations and interpolation to assign dose on the original CT grid. Being optimized for GPU acceleration, DoTA is the fastest alternative, needing less than 15 seconds to calculate full dose distributions. For the baselines in this paper, we find that PBA runtimes oscillate between 100 and 150 seconds, while B1 and B2 report needing only few seconds to correct/denoise their inputs, but must add the runtime necessary to generate their respective PBA (\SIrange{123}{303}{\second} in \cite{Wu2021}) or MC ($\approx\SI{10}{\second}$ in \cite{Javaid2021})) input doses, as well as data transfer times between the physics engine and the deep learning framework. Furthermore, B2 is a per beam network, hence its runtime scales linearly with the number of beams, in practice meaning 2-4 times higher total calculation times.

\begin{table}
\begin{minipage}{0.48\linewidth}
	\caption{\textbf{Beamlet prediction runtime}. The reported values include the mean inference time and standard deviation (Std) taken by each model to predict individual beamlet dose distributions. Both the DoTA and LSTM models run on GPU hardware, while the pencil beam algorithm (PBA) and Monte Carlo (MC) dose engine use CPUs with multiple threads. LSTM inference times are taken directly from \cite{Neishabouri2021}. }
	\begin{center}
		\begin{tabular}{@{}lcc@{}}
			\toprule
			%\multicolumn{3}{c}{Inference time} \\
			\textbf{Model} & \textbf{Mean (ms)} & \textbf{Std (ms)} \\ \midrule
			%LSTM$^a$ & 23.0 & 2.5 \\
			LSTM$^a$\cite{Neishabouri2021} & 6.0 & 1.5 \\
			DoTA$^b$ & 5.0 & 4.9 \\
			PBA$^c$\cite{Wieser2017} & 728.3 & 30.9 \\
			MC$^c$\cite{Souris2016} & 43,636.9 & 12,291.6 \\ \bottomrule
		\end{tabular}\\
	\end{center}				
	\label{tab:time}
	\footnotesize{$^a$ Nvidia{\small\textregistered} Quadro RTX 6000 64 Gb RAM}\\
	\footnotesize{$^b$ Debian 10 4 vCPUs - Nvidia{\small\textregistered} A100 40 Gb RAM}\\
	\footnotesize{$^c$ CentOS 7 8 CPUs intel Xeon{\small\textregistered}} E5-2620 2 GHz 16Gb RAM
\end{minipage}\hfill
\begin{minipage}{0.48\linewidth}
	\centering
	\includegraphics[width=\textwidth]{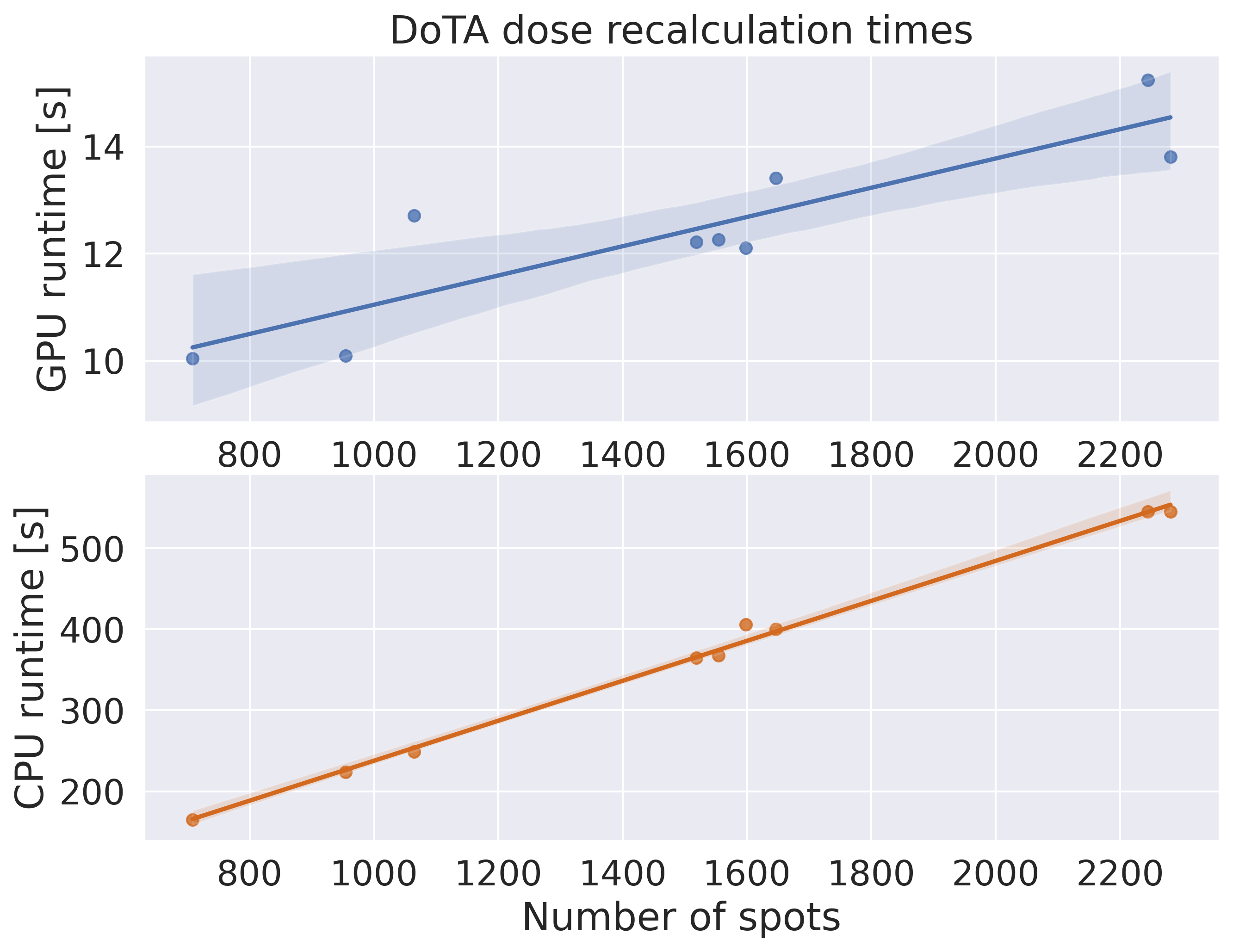}
	\captionof{figure}{\textbf{Full dose recalculation runtime}. Time needed to recalculate planned dose distributions with DoTA using (top) a Nvidia{\small\textregistered} A100 GPU or (bottom) an intel Xeon{\small\textregistered} CPU. Estimates include time for loading CT and beam weights from plan data, for dose inference by DoTA and for the necessary CT and dose interpolations. Shaded areas denote the 95\% confidence interval.}
	\label{timeplans}
\end{minipage}
\end{table}

\section{Discussion}
\label{sec:Discussion}
% Individual beamlets and dose recalculation.
In this study, we present a data-driven dose engine predicting dose distributions with high accuracy. The presented DoTA model builds upon previous work learning proton transport as sequence modeling task via LSTM networks \cite{Neishabouri2021}, by introducing energy dependence and significantly improving its performance in a varied set of treatment sites. DoTA greatly outperforms analytical physics-based PBA algorithms in predicting dose distributions from individual proton pencil beams, achieving high accuracy even in the most heterogeneous patient geometries, demonstrated by the 6\% improvement in the minimum gamma pass rate. With millisecond inference times, DoTA provides at least a factor 100 reduction in calculation time compared to the clinically still predominant analytical PBAs. 

The drastic reduction in spot dose prediction times translates into the ability to calculate full dose distributions in \SI{12}{\second} on average and less than \SI{15}{\second} even for the plan with more than 2200 pencil beams, which times include the required time for all steps from loading CT and pencil beam weights from plan data ($\approx\SI{1}{\second}$ on average), CT interpolation and beamlet geometry extraction ($\approx\SI{1}{\second}$), DoTA model and weights loading ($\approx\SI{2}{\second}$), dose inference by DoTA ($\approx\SI{7.5}{\second}$) and interpolating the final dose distribution back to the original CT grid ($\approx\SI{1}{\second}$). We achieve this \SIrange{10}{15}{\second} speed on a single GPU, even without any optimization of GPU settings for inference, which can reportedly yield up to 9 times speed-ups depending on the task \footnote{Discussed in the non-peer-reviewed study in \url{https://huggingface.co/transformers/v2.10.0/benchmarks.html}}. Without sacrificing accuracy, DoTA represents at least a factor 10 speed-up with respect to PBAs and a 33\% speed-up (and $\approx$ 80\% considering the difference in MC noise levels) with respect to the fastest GPU MC competitor we could find in the literature --- clinically used GPU MC software Raystation{\textregistered} \cite{Fracchiolla2021}, typically running in clusters or workstations with multiple GPUs and CPU cores. Moreover, DoTA offers a 10-25\% increase in the $\Gamma(1\text{ mm},1\%)$ gamma pass rate compared to PBA, and with a $\Gamma(2\text{ mm},2\%)$ gamma pass rate >99\% it matches \cite{Wang2016} or outperforms \cite{Tseung2015,Qin2016} the accuracy of GPU MC approaches. DoTA's accuracy is also on par with the agreement between commercial MC engines (Raystation{\textregistered}) and experimental measurements \cite{Schreuder2019a, Schreuder2019b}. While the GPU-based PBA algorithm reported in \cite{Silva2015} calculates a full distribution in $\SI{0.22}{\second}$ and is faster than DoTA, it was tested only on a single patient showing worse accuracy with a 3\% lower $\Gamma(2\text{ mm},2\%)$ pass rate.

Our method is also substantially superior to the only 2 published deep learning approaches for proton full plan dose calculations \cite{Javaid2021,Wu2021}. We achieve 15\% and 25\% higher $\Gamma(2\text{ mm},2\%)$ pass rates compared to the MC-denoising U-net of \cite{Javaid2021}, and 6\% and 2\% higher $\Gamma(1\text{ mm},1\%)$ pass rates compared to the PBA correcting U-net of \cite{Wu2021} in lung and H\&N patients, respectively. DoTA shows a slight inferiority in prostate patients, with a $\approx 3\%$ lower $\Gamma(1\text{ mm},1\%)$ pass rates than \cite{Wu2021}. However, this direct comparison is somewhat unfair to DoTA. In \cite{Wu2021}, double scattering proton therapy plans were used, while in our work we evaluate performance on Intensity Modulated Proton Therapy plans with a small, $\SIrange{3}{5}{\milli\meter}$ spot size, making our plans significantly more conformal, less smooth and more sensitive, translating into a more difficult dose calculation task. We also use a finer voxel resolution of $\SI{2}{\milli\meter}\times\SI{2}{\milli\meter}\times\SI{2}{\milli\meter}$ compared to the $\SI{2}{\milli\meter}\times\SI{2}{\milli\meter}\times\SI{2.5}{\milli\meter}$ used in \cite{Wu2021}. Furthermore, \cite{Wu2021} also reports site specific fine-tuning of their deep learning approach, unlike our method. Last, \cite{Wu2021} has the further disadvantage of using per beam PBA calculations as input, thus the reported $\SIrange{2}{3}{\second}$ dose correction times easily translate to full treatment plan calculation times in the \SIrange{5}{10}{\minute} range depending on the number of beams (taking into account the >\SI{2}{\minute} PBA run times), even without accounting for the additional time for the necessary CT rotations and interpolations.

DoTA's accuracy may further be increased by training with larger datasets, as demonstrated by the improvement achieved when increasing training data from 4 lung patients in our earlier work \cite{Pastor-Serrano2021} to 30 patients with varied anatomies in the current study. Using dose distributions with lower MC noise could further improve performance. Convincingly outperforming all recent works learning corrections for 'cheap' physics-based predictions \cite{Wu2021, Javaid2021} both in terms of accuracy and speed, DoTA has the flexibility to be used in a great variety of treatment sites and clinical settings. 

\paragraph{Application}
% From x to breathing interplay effects evaluation whose clinical implementation is currently limited by the computational requirements of calculating the dose from thousands of pencil beams in many breathing scenarios.
DoTA's accuracy and speed improvements outperform existing approaches and represent a new state-of-the-art that could benefit current RT practice in numerous aspects. The small number of potential geometries currently used to evaluate treatment plan robustness --- whose size is limited by the speed of the dose calculation algorithm --- can be extended with many additional samples, capturing a more diverse and realistic set of inter- and intra-fraction \cite{PastorSerrano2021} geometrical variations. DoTA's capability to quickly and accurately estimate fraction dose distributions based on pre-treatment daily CT images could transform dosimetric quality assurance protocols, enabling direct comparison between the planned and estimated doses or even online adaptation of plans \cite{Jagt2017, Jagt2018, Albertini2020}. Most crucially, by pre-computing the input volumes and updating their CT values in real time, the millisecond speed for individual pencil beam dose calculation makes our model well suited for real-time correction during radiation delivery. 

\paragraph{Limitations} The current version of DoTA is trained to predict MC ground truth dose distributions from a specific machine with unique settings and beam profiles, necessitating a specific model per machine. Likewise, range shifters --- which are often dependent on treatment location and site --- affect the dose delivered by some spots while inserted, thereby modifying the final dose distribution. Both problems could in principle be addressed by constructing a model that takes extra shape and range shifter specifications as input in the form of tokens at the beginning of the sequence, similar to our approach for treating the energy dependence.

Moreover, DoTA is trained for a specific voxel grid resolution, requiring either an individual model per resolution level or an additional interpolation step that will likely negatively interfere with the gamma pass rate results, especially for gamma evaluations $\Gamma(1,1\%)$ with a distance-to-agreement criterion lower than the voxel resolution level. While DoTA also works for finer nominal CT grids \cite{Pastor-Serrano2021}, an additional study testing the dose recalculation performance with more patients and finer grid resolution should confirm its suitability for direct clinical application needing such resolutions.

\paragraph{Future work}
Besides the possibility to include shape, machine and beam characteristics as additional input tokens in the transformer, several extensions can widen its spectrum of applications, such as predicting additional quantities, e.g., particle flux, or estimating radiobiological weighted dose -- potentially including simulating even DNA damage -- typically significantly slower than pure MC dose calculation. Alternatively, future work adapting DoTA to learn photon physics would facilitate its use in conventional radiotherapy applications or provide CT/CBCT imaging reconstruction techniques with the necessary speed for real-time adaptation. Most importantly, DoTA offers great potential to speed up dose calculation times in heavy ion treatments with particles such as carbon and helium sharing similar, mostly forward scatter physics, whose MC dose calculation often take much longer to simulate all secondary particles generated as the beam travels through the patient.

\section{Conclusion}
We present DoTA: a generic, fast and accurate dose engine that implicitly learns proton particle transport physics and can be applied to speed up several steps of the radiotherapy workflow. Framing particle transport as sequence modeling of 2D geometry slices in the proton's beam travel direction, we use the power of transformers to predict individual beamlets with millisecond speed and close to MC precision. Our evaluation shows that DoTA has the right attributes to potentially replace the proton dose calculation tools currently used in the clinics for applications that critically depend on runtime. Predicting dose distributions from single pencil beams in milliseconds, DoTA offers 100 times faster inference times than widely used PBAs, yielding close to MC accuracy as indicated by the very high gamma pass rate $\Gamma(3\text{ mm},1\%)$ of $99.37\pm1.17$, thus has the potential to enable next generation online and real-time adaptive radiotherapy cancer treatments. The presented model predicts MC quality full plan dose distributions with at least a 10\% improvement in gamma pass rate $\Gamma(1\text{ mm},1\%)$ with respect to current analytical approaches and reduces dose calculation times of planned doses to less than 15 seconds, representing a tool that can directly benefit current clinical practice too.

\section*{Acknowledgments}
This work is supported by KWF Kanker Bestrijding [grant number 11711] and is part of the KWF research project PAREL. Zolt\'an Perk\'o would like to thank the support of the NWO VENI grant ALLEGRO (016.Veni.198.055) during the time of this study. 

\section*{Code availability}
The code, weights and results are publicly available at \url{https://github.com/}.

\section*{CRediT authorship contribution statement}
\textbf{Oscar Pastor-Serrano}: Conceptualization, Methodology, Software, Validation, Formal Analysis, Investigation, Data Curation, Writing – original draft, Visualization.

\noindent \textbf{Zolt\'an Perk\'o}: Conceptualization, Methodology, Formal Analysis, Resources, Writing – original draft, Writing – Review \& editing, Supervision, Project Administration, Funding Acquisition.

\small
\bibliographystyle{bibstyle}
\bibliography{DoTA_journal}

\normalsize
\appendix
\section{Transformer and self-attention}
\label{Appendix:attention}
\paragraph{Transformer} DoTA's backbone is the Transformer \cite{Vaswani2017}, based on the self-attention mechanism. Though originally introduced for sequential modeling applications in natural language processing such as machine translation, Transformers have recently achieved state-of-the-art performance across a wide variety of tasks, with large language \cite{Devlin2019, Brown2020} or computer vision \cite{Dosovitskiy2020} models replacing and outperforming recurrent or convolutional architectures. One of the main reasons behind the success of attention-based models is the ability to model interactions between a large sequence of elements without needing an internal memory state. In Transformers, each sequence element is transformed based on the information it selectively gathers from other members of the sequence based on its content or position. In practice, however, the computational memory requirements scale quadratically with the length of the sequence, and training such large Transformers often requires a pre-training stage with a large amount of data.

\paragraph{Self-attention} Given a sequence $\bm{z}\in\mathbb{R}^{L\times D}$ with $L$ tokens, the self-attention (SA) mechanism \cite{Vaswani2017} is based on the interaction between a series of queries $\bm{Q}\in\mathbb{R}^{L\times D_h}$, keys $\bm{K}\in\mathbb{R}^{L\times D_h}$, and values $\bm{V}\in\mathbb{R}^{L\times D_h}$ of dimensionality $D_h$ obtained through a learned linear transformation of the input tokens with weights $\bm{W}_{QKV}\in\mathbb{R}^{D\times 3D_h}$ as

\begin{equation}
	[\bm{Q}, \bm{K}, \bm{V}] = \bm{z}\bm{W}_{QKV}.
\end{equation}

Each token is transformer into a query, key and value vector. Intuitively, for an $i^{th}$ token $\bm{z}_i\in\mathbb{R}^{1\times D}$, the query $\bm{q}_i\in\mathbb{R}^{1\times D_h}$ represents the information to be gathered from other elements of the sequence, while the key $\bm{k}_i\in\mathbb{R}^{1\times D_h}$ contains token's information to be shared with other sequence members. The token $\bm{z}_i$ is then transformed into $\bm{z}_i'$ via a weighted sum of all values in the sequence $\bm{v}_j\in\mathbb{R}^{1\times D_h}$ as

\begin{equation}
	\bm{z}_i' = \sum_{j=1}^{L} w_j \bm{v}_j,
\end{equation}

\noindent where each weight is based on a the similarity between the $i^{th}$ query and the other keys in the sequence, measured as the dot product $w_j = \bm{q}_i^T\bm{k}_j$. The output sequence of transformed tokens $\bm{z}\in\mathbb{R}^{L\times D}$ is the result of the SA operation applied to all sequence elements, defined by the attention matrix containing all weights $\bm{A}\in\mathbb{R}^{L\times L}$ and the operations

\begin{equation}
	\bm{A} = \text{softmax}\Big(\frac{\bm{Q}\bm{K}^T}{\sqrt{D_h}}\Big),
\end{equation}

\begin{equation}
	\bm{z}' = \text{SA}(\bm{z}) = \bm{A}\bm{V}.
\end{equation} 

A variant of SA called multi-head self-attention (MSA) runs $N_h$ parallel SA operations focusing on different features or inter-dependencies of the data. Setting $D_h=D$, the outputs of the different SA operations, called \textit{heads}, are first concatenated and then linearly projected with learned weights $\bm{W}_h\in\mathbb{R}^{N_hD_h\times D}$ as

\begin{equation}
	\text{MSA}({\bm{z}}) = \underset{h\in\{N_h\}}{\text{concat}}[\text{SA}_h(\bm{z})]\bm{W}_h.
\end{equation}

By definition, every token can attend to all previous and future tokens. Causal SA is a variant of SA applied to sequence modeling tasks restricting access to future information, where all elements above the diagonal in the attention matrix $\bm{A}$ are masked to 0. Additionally, since SA is invariant to the relative order of elements in the sequence, a fixed \cite{Vaswani2017} or learned \cite{Dosovitskiy2020} positional embedding $\bm{r}\in\mathbb{R}^{L\times D}$ is usually added or concatenated to the input tokens, where is element in the positional embedding sequence contains unique information about its position.

\paragraph{Transformer encoder} The causal MSA Transformer backbone in DoTA is responsible of routing information between the geometry slices and the energy token. A learnable positional embedding  $\bm{r}$ is added to the sequence of tokens produced by the convolutional encoder, while we add the first 0th position embedding $\bm{r}_0$ in the sequence to the energy token. The transformer encoder is formed by alternating MSA and Multi-layer Perceptron (MLP) layers with residual connections, and applying Layer Normalization (LN) applied before every layer \cite{Ba2016}. Therefore, the Transformer encoder blocks computes the operations

\begin{equation}
	\bm{z} = [\bm{z}_e;\bm{z}] + \bm{r},
\end{equation}
\begin{equation}
	\bm{s}_n = \bm{z} + \text{MSA}(\text{LN}(\bm{z})),   
\end{equation}
\begin{equation}
	\bm{z}' = \bm{s}_{n} + \text{MLP}(\text{LN}(\bm{s}_{n})),
\end{equation}

\noindent where MLP denotes a two layer feed-forward network with Dropout \cite{Srivastava2014} and Gaussian Error Linear Unit (GELU) activations \cite{Hendrycks2016}.

\section{Gamma analysis}
\label{Appendix:gamma}
The gamma analysis is based on the notion that doses delivered in neighboring voxels have similar biological effects. Intuitively, for a set reference points --- the voxel centers in the ground truth 3D volume --- and their corresponding dose values, this method searches for similar predicted doses within small spheres around each point. The sphere's radius is referred to as distance-to-agreement criterion, while the dose similarity is usually quantified as a percentage of the reference dose, e.g., dose values are accepted similar if within 1\% of the reference dose. Each voxel with coordinates $\bm{a}$ in the reference grid is compared to points $\bm{b}$ of the predicted dose grid and assigned a gamma value $\gamma(\bm{a})$ according to

\begin{equation}
	\gamma(\bm{a}) = \underset{\bm{b}}{\min}\{\Gamma_{\bm{a},\bm{b}}(\delta, \Delta)\},
\end{equation}

\begin{equation}
	\Gamma_{\bm{a},\bm{b}}(\delta, \Delta)=\sqrt{\frac{\abs{\bm{a}-\bm{b}}^2}{\delta^2}+\frac{\abs{\hat{y}_{\bm{a}}-y_{\bm{b}}}^2}{\Delta^2}},
\end{equation}

\noindent where $\hat{y}_{\bm{a}}$ is the reference dose at point $\bm{a}$, $\delta$ is the distance-to-agreement, and $\Delta$ is the dose difference criterion. A voxel passes the gamma analysis if $\gamma(\bm{a})<1$. 

\end{document}